\documentclass[aps,prd,reprint,superscriptaddress]{revtex4-2}

\usepackage{graphicx}
\usepackage{amsmath}
\usepackage{siunitx}
\usepackage{xcolor}
\usepackage{booktabs}

\begin{document}

\title{A high-resolution search for axion dark-matter flows with ORGAN-Q}
\author{Michael T. Hatzon}
\email{michael.hatzon@uwa.edu.au}
\affiliation{Quantum Technologies and Dark Matter Laboratory,
Department of Physics, University of Western Australia,
Crawley, WA 6009, Australia}

\author{Aaron P. Quiskamp}
\affiliation{Quantum Technologies and Dark Matter Laboratory,
Department of Physics, University of Western Australia,
Crawley, WA 6009, Australia}

\author{Graeme R. Flower}
\affiliation{Quantum Technologies and Dark Matter Laboratory,
Department of Physics, University of Western Australia,
Crawley, WA 6009, Australia}

\author{Steven Samuels}
\affiliation{Quantum Technologies and Dark Matter Laboratory,
Department of Physics, University of Western Australia,
Crawley, WA 6009, Australia}

\author{Ben T. McAllister}
\affiliation{Quantum Technologies and Dark Matter Laboratory,
Department of Physics, University of Western Australia,
Crawley, WA 6009, Australia}
\affiliation{Centre for Astrophysics and Supercomputing, Swinburne University of Technology, Hawthorn VIC 3122, Australia}

\author{Paul Altin}
\affiliation{Department of Quantum Science, Research School of Physics, The Australian National University, Canberra ACT 2600, Australia}

\author{Eugene N. Ivanov}
\affiliation{Quantum Technologies and Dark Matter Laboratory,
Department of Physics, University of Western Australia,
Crawley, WA 6009, Australia}

\author{Maxim Goryachev}
\affiliation{Quantum Technologies and Dark Matter Laboratory,
Department of Physics, University of Western Australia,
Crawley, WA 6009, Australia}

\author{Michael E. Tobar}
\affiliation{Quantum Technologies and Dark Matter Laboratory,
Department of Physics, University of Western Australia,
Crawley, WA 6009, Australia}

\date{\today}

\begin{abstract}
    We search for narrow axion dark-matter flows using the high-resolution data acquired during the ORGAN-Q experiment. The analysis covers nine potential axion linewidths from \SI{0.466}{Hz} to \SI{119.209}{Hz}, equivalent to axion effective quality factors $1.3\times 10^{10}$ to $5.2\times 10^7$, over the range \SIrange{6.1550}{6.3496}{GHz}, corresponding to axion masses of approximately \SIrange{25.455}{26.260}{\micro \electronvolt}. Once contributions from native-bin correlations, data-quality and interference vetoes, frequency-dependent system noise, the cavity response, and filter attenuation are accounted for, no candidate satisfies the required persistence ratio in any search bandwidth. Across the nine assumed linewidths, the strongest $95\%$ KSVZ axion-flow density limits range from \SI{22.8}{\mathrm{GeVcm^{-3}}} for a $\SI{0.466}{Hz}$ signal to \SI{153.7}{GeVcm^{-3}} for a $\SI{119.209}{Hz}$ signal, with both attained near $m_a=\SI{25.857}{\micro\electronvolt}$.
\end{abstract}

\maketitle

\section{Introduction}
\label{sec:intro}
The axion was originally introduced as a consequence of the Peccei--Quinn mechanism, which provides a dynamical solution to the strong CP problem of quantum chromodynamics \cite{peccei1977cp,weinberg1978new,wilczek1978problem}. In addition to this particle-physics motivation, axions produced non-thermally in the early universe can behave as cold dark matter and may constitute some or all of the Galactic dark-matter density \cite{preskill1983cosmology,abbott1983cosmological,dine1983not}. Their small mass and weak interactions make direct detection challenging, but the axion-photon coupling provides a measurable electromagnetic signature.

In a microwave-cavity haloscope, Galactic axions may convert into photons in the presence of a strong static magnetic field \cite{PhysRevLett.51.1415,sikivie1985detection}. When a cavity mode is tuned near the axion conversion frequency, the deposited power is resonantly enhanced and can be detected through the receiver chain. Conservation of energy relates the frequency $\nu_a$ of the resulting photon to the energy of a non-relativistic axion according to \begin{equation}
h \nu_a \simeq m_a c^2 +\frac{1}{2}m_a v^2, 
\label{eq:axion_energy}
\end{equation} where $h$ is Planck's constant, $m_a$ is the axion mass, $c$ is the speed of light, and $v$ is the axion velocity relative to the detector. For an axion flow with velocity dispersion $\Delta v$, this relation gives a fractional signal linewidth \begin{equation}
    \frac{\Delta \nu_a}{\nu_{a,0}}\simeq \frac{v \Delta v}{c^2},
\end{equation} where $\Delta \nu_a$ is the signal linewidth, $\nu_{a,0}=m_a c^2/h$ is the axion rest-mass frequency, and $\Delta v$ is the velocity dispersion of the flow. The central signal frequency therefore primarily identifies the mass, while the signal linewidth and temporal frequency variation contain information about the local axion velocity distribution.

The Oscillating Resonant Group AxioN (ORGAN) experiment is a microwave-cavity haloscope designed primarily to search for axions at comparatively high masses \cite{mcallister2017organ}. The ORGAN-Q experiment used a tunable conducting-rod resonator, a strong solenoidal magnetic field, and operated at milli-kelvin with a Josephson parametric amplifier to search the frequency interval from approximately $\SIrange{6.16}{6.35}{GHz}$. It provided constraints on the virialised axion mass range over $\SIrange{25.45}{26.27}{\mu \mathrm{eV}}$ \cite{quiskamp2025near}.

Conventional haloscope analyses generally adopt a smooth, approximately virialised Galactic halo, for which characteristic velocities of order $10^{-3}c$ produce a fractional axion linewidth of order $10^{-6}$ \cite{PhysRevD.42.3572,PhysRevD.96.123008,quiskamp2025near}. At the ORGAN-Q operating frequency, this corresponds to a signal distributed over several kilohertz. However, galactic dark matter may consist of more coherent populations. Fine-grained substructure may produce flows with much smaller velocity dispersions and correspondingly narrower spectral features. Such models include cold infall and caustic ring flows, together with streams produced by the tidal disruption of axion miniclusters \cite{sikivie1998caustic,PhysRevD.93.103509,tinyakov2016tidal}. The assumptions of a caustic ring model for example, produce an upper limit to the fractional energy dispersion of the ``Big Flow'' dark matter stream of $\delta E / (m_a c^2) \lesssim 2.4 \times 10^{-10}$ \cite{PhysRevD.93.103509}, with $\delta E$ the root-mean-square energy dispersion of the flow. This corresponds to a lower limit of effective quality factor \(Q_{\mathrm{rms}}\equiv m_a c^2/\delta E\gtrsim4\times10^9\) or an upper limit on the signal width on the order of 1.5~Hz at ORGAN-Q frequencies. This model also provides an axion density central estimate of \SI{11.2}{GeV/cm^3} \cite{CHAKRABARTY2021100838}, providing a model-dependent benchmark for interpreting limits on narrow flows. High-resolution cavity searches have previously been used to probe such non-virialised axion populations \cite{PhysRevD.74.012006,PhysRevD.84.121302,PhysRevD.94.082001,PhysRevD.109.083014}. 

The ORGAN-Q acquisition system generated two simultaneous data streams: a medium resolution stream, previously used to report on a virialised halo search, and a high-resolution stream, used here to search for narrower signals (see Sec.~\ref{sec_high_res} for details). Motivated by these possibilities, we use the ORGAN-Q high-resolution data to search for persistent narrow spectral features and constrain the density of non-virialised KSVZ axion flows.

\section{Experiment and Data Set}

\subsection{ORGAN apparatus and operating configuration}
The ORGAN-Q experiment ~\cite{quiskamp2025near} employed a $\SI{0.708}{L}$ oxygen-free high-conductivity copper cavity inside a $\SI{7}{T}$ solenoidal magnet, for which the average magnetic field over the cavity volume was $B_0=\SI{6.26\pm0.01}{T}$. An off-axis aluminium tuning rod varied the frequency of the TM$_{010}$ mode between approximately $\SIrange{6.15}{6.35}{GHz}$ without encountering a mode crossing, yielding typical mode form factors of $C\simeq0.42$. Over the rod-tuning range of the experiment, the loaded quality factor was $Q_L\simeq 1700$ ($\sigma =400)$, whilst the antenna coupling had an average value of $\beta \simeq 1.45$ ($\sigma=0.32$).

\begin{figure*}
    \centering
    \includegraphics[width=0.8\linewidth]{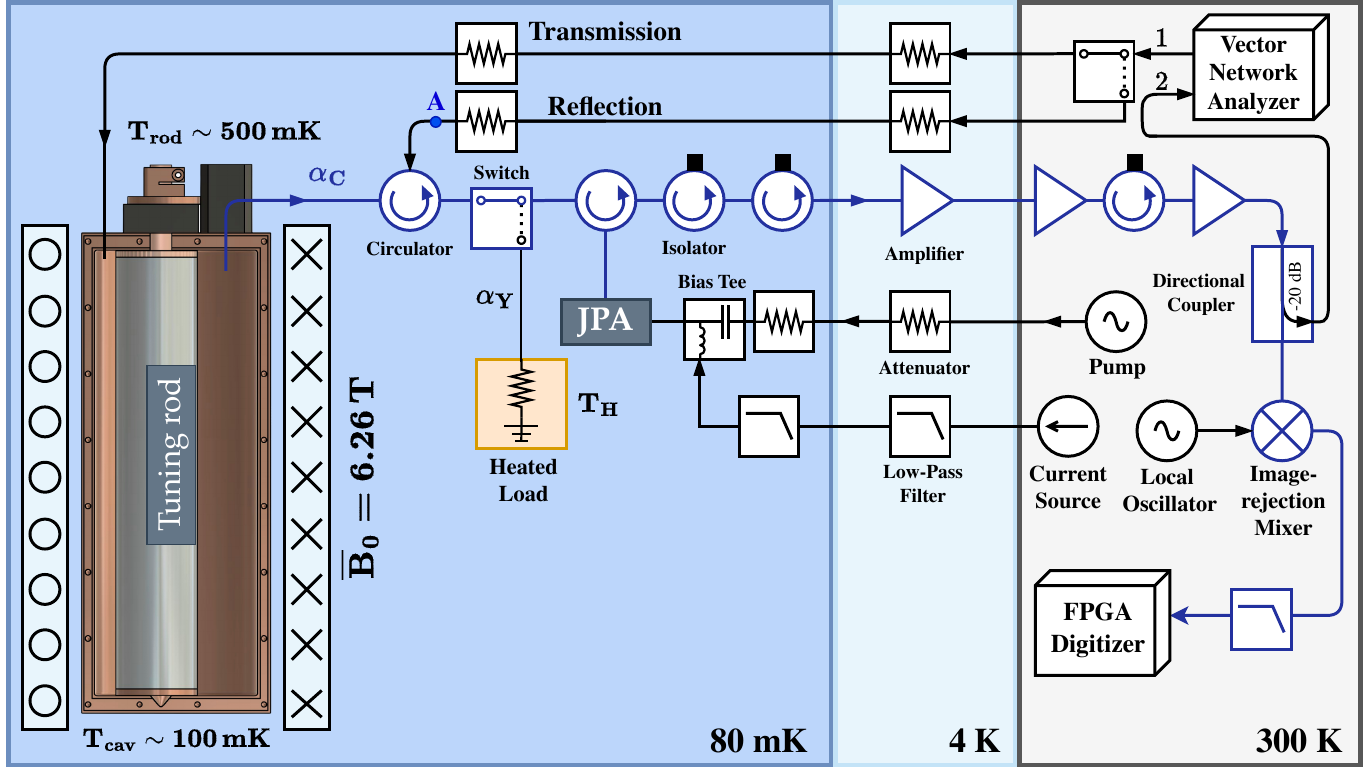}
    \caption{Schematic of the ORGAN-Q apparatus and receiver chain used to acquire the high-resolution data. A potential axion signal generated in the tunable cavity is routed through the cryogenic receiver to the flux-driven JPA and subsequent HEMT amplification stage. At room temperature, the amplified signal is down-converted using an image-rejecting mixer before being processed by the FPGA digitiser. The vector network analyser, cryogenic switch and heated load provide the measurements required for a cavity and receiver calibration. Piezoelectric actuators control the tuning-rod position and antenna coupling. The schematic is reproduced from Ref.~\cite{quiskamp2025near}.}
    \label{fig:setup}
\end{figure*}

The experimental configuration is shown schematically in Fig.~\ref{fig:setup}. The cavity and first-stage receiver components operated at millikelvin temperatures. The mixing chamber and cavity were maintained near $\SI{80}{mK}$ and $\SI{100}{mK}$, respectively, although motion of the tuning rod produced a higher characteristic rod temperature of approximately $\SI{500}{mK}$ \cite{quiskamp2025near}. Signals from the strongly coupled cavity port were routed through a cryogenic circulator to a flux-driven Josephson parametric amplifier (JPA), followed by a cryogenic high-electron-mobility-transistor (HEMT) amplifier. The JPA was operated in three-wave-mixing-mode, with a pump frequency near twice the signal frequency, and provided more than $\SI{18}{dB}$ of gain over most of the searched range. Following amplification, the signal was mixed to an intermediate frequency using an in-phase and quadrature mixer, whose outputs were combined using a $90^\circ$ hybrid to provide image rejection before digitisation.

At each tuning position, vector-network-analyser measurements tracked the cavity resonance and JPA gain response. The tuning increment was approximately one fifth of the cavity linewidth, and acquisition at an individual tuning position typically lasted between $\SIrange{30}{60}{min}$. The medium- and high-resolution data streams were acquired simultaneously by the field-programmable gate array (FPGA) digitiser.

\subsection{Data acquisition}
\label{sec_high_res}
The medium-resolution stream performed real-time averaging and produced $6554$-bin spectra over a $\SI{6.25}{MHz}$ interval centred near an intermediate frequency of $\SI{45}{MHz},$ corresponding to a bin spacing of $\approx\SI{954}{Hz}$. These data also supplied the tuning and receiver calibration metadata used in the present analysis. In parallel, the high-resolution stream operated at one quarter of the primary sampling rate and without real-time spectral averaging. For the high-resolution stream, time-series data were transferred continuously, without gaps, from the FPGA to the host computer, where FFTs were computed to produce the stored power spectra. Each stored high-resolution spectrum contains $3,355,444$ bins spanning $\SI{1.5625}{MHz}$ with frequency spacing $\delta f \approx \SI{0.466}{Hz}.$ However, after analysing the data, frequency-bin correlations were detected. These are consistent with rectangular records zero-padded by a representative oversampling factor $R\approx 4.272$, as discussed and accounted for in Appendix~\ref{app:fft_response}. This corresponds to a representative record duration $T_{\mathrm{rec}}=1/(R \delta f)\approx0.503$~s, and a rectangular-window equivalent noise bandwidth $1/T_{\mathrm{rec}}\approx 1.99$~Hz.

Two passes over the frequency range were acquired from 13--26 February 2024 and 6--10 March 2024. The second pass over selected frequency regions were to provide sufficient coverage after accounting for interference or poor receiver performance.

\subsection{System-noise calibration}
\label{sec:tsys}
The frequency-dependent system-noise temperature referred to the cavity output $T_{\mathrm{sys},i}$, was determined using the signal-to-noise ratio improvement (SNRI) procedure discussed in Ref.~\cite{quiskamp2025near} for each tuning position. This combined the measured HEMT noise temperature and receiver-line attenuation with JPA-off transfer functions measured using a vector network analyser and noise-power spectra recorded using a spectrum analyser. The same $T_{\mathrm{sys},i}$ cuts and calibration employed in the medium-resolution were used for this search, satisfying $0.4~K<T_{\mathrm{sys},i}<1.35~K$.

\section{Analysis Pipeline}
\label{sec:pipeline}

\subsection{Baseline Filtering}
\begin{figure*}[htbp]
    \centering
    \includegraphics[width=0.99\linewidth]{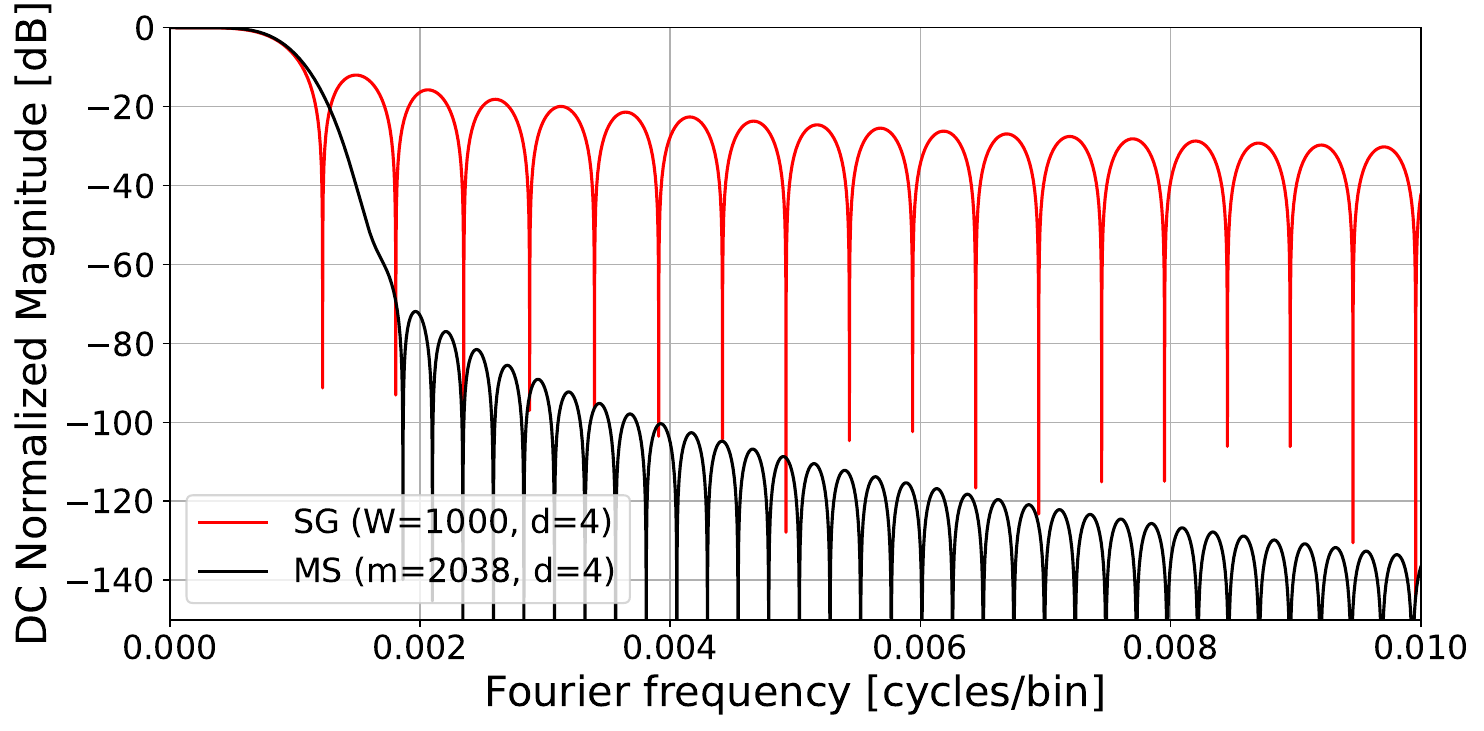}
    \caption{Representative magnitude responses of degree-four Savitzky--Golay (SG) and modified-sinc (MS) smoothers matched at their $-3$-dB cutoff. The responses are normalised at zero spatial frequency, expressed as cycles per native spectral bin. The MS kernel exhibits substantially lower stop-band sidelobes.}
    \label{fig:msvssg}
\end{figure*}

Each raw high-resolution spectrum may contain smooth frequency-dependent structure arising from the receiving gain and noise response. This baseline must be estimated and divided out before the statistical properties of the residual power can be tested. Savitzky--Golay (SG) filters are commonly used for this purpose in haloscope analyses. In the present analysis, the baseline was instead estimated using the modified-sinc (MS) smoother introduced in Ref.~\cite{schmid2022and}. At a matched $3$-dB cutoff, the MS filter provides substantially stronger stop-band suppression than a conventional SG filter \cite{schmid2022and}. The transfer functions of the two filters at a demonstrative window are shown in Fig.~\ref{fig:msvssg}. Despite stronger stop-band suppression, this alone does not establish improved recovery of narrow spectral signals. Such signals have broad Fourier spectra, and their attenuation after baseline normalisation depends on the full filter response and the signal profile. The MS smoother was therefore selected for its strong suppression of fine-scale structure in the estimated baseline, with its signal-recovery evaluated using injection tests described in Sec.~\ref{Sec:filter_attenuation}.

Denoting the linear power in native frequency bin $i$ by $P_i$, the estimated baseline was \begin{equation}
    \hat{b}_i=\sum_{j=-m}^m h^{MS}_j P_{i-j},
\end{equation} where $\hat{b}_i$ is the estimated baseline in bin $i$, $h^{MS}_j$ is the normalised symmetric MS kernel at bin offset $j$, and $m$ is the kernel half-width. The analysis used the degree parameter $d=4$ and $m=100,000$, corresponding to a $2m+1$ bin kernel. At the boundaries of each spectrum, the data required by the convolution were extrapolated using cosine-squared-weighted linear extrapolation. The convolution was evaluated using zero-padded FFT convolution to assist with edge-handling.

Bins belonging to manually identified strong-features masks discussed in Sec.~\ref{section:cuts}, had their power replaced by the mean power measured ten masked-interval widths either side of the mask. The masked bins were omitted from the Erlang test in Sec.~\ref{section:erlang} and excluded from the subsequent trigger search.

\subsection{Noise Distribution Cuts}
\label{section:erlang}
\begin{figure}[htbp]
    \centering
    \includegraphics[width=0.99\linewidth]{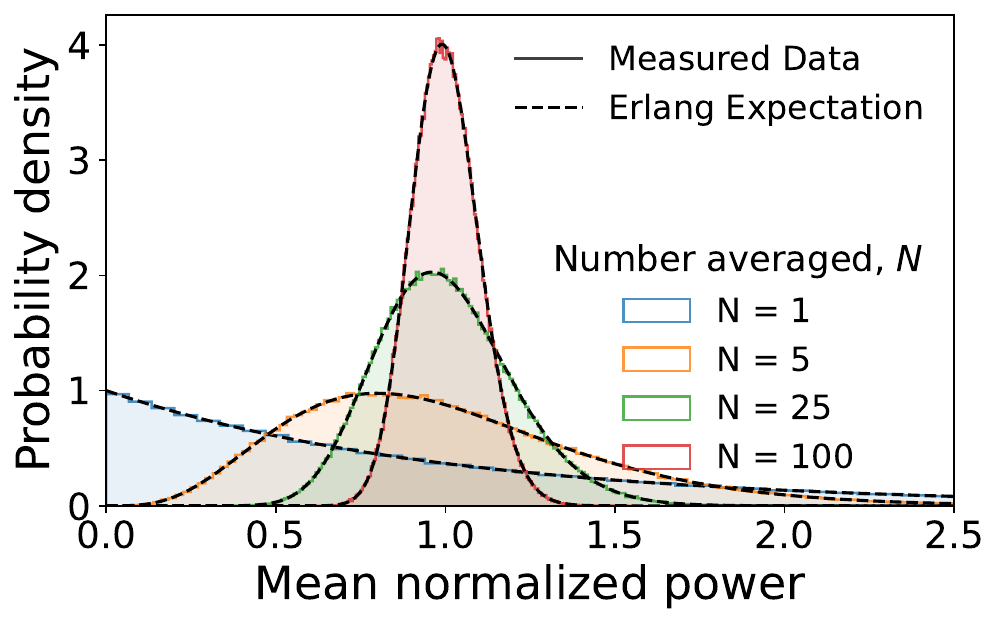}
    \caption{Erlang evolution expectation with comparison to the data.}
    \label{fig:erlang_evolution}
\end{figure}

Following baseline normalisation, the power distribution of each raw spectrum was tested for consistency with an Erlang distribution;
\begin{equation}
    p(x;k,\beta)
    =
    \frac{x^{k-1}e^{-x/\beta}}{\beta^k (k-1)!} ,
    \qquad x \geq 0 ,
    \label{erlang}
\end{equation}

where $x$ is the baseline-normalised power, $k$ is the shape parameter and $\beta$  is the scale parameter. For an individual, unaveraged spectrum, $k=1$ and $\beta$ is its mean power, reducing Eq.~\ref{erlang} to the exponential distribution expected for stationary Gaussian receiver noise. The evolution of the erlang distribution is shown in Fig.~\ref{fig:erlang_evolution} and compared to a sample dataset.

The agreement between the measured distribution of a spectrum and the expected Erlang distribution is quantified using the reduced Pearson's chi-square statistic \cite{pearson1900x},

\begin{equation}
    \chi^2_\nu
    =
    \frac{1}{\nu}
    \sum_i
    \frac{(O_i-E_i)^2}{E_i},
\end{equation}
where $O_i$ and $E_i$ are the observed and expected counts in histogram bin i, respectively, and $\nu$ is the degrees of freedom. Every tenth native frequency bin was sampled, known strong signals masked, constructed using 1000 equal-width bins with those $E_i <5$ excluded. The expected counts were normalised to the corresponding observed total.

\begin{figure}[htbp]
    \centering
    \includegraphics[width=0.99\linewidth]{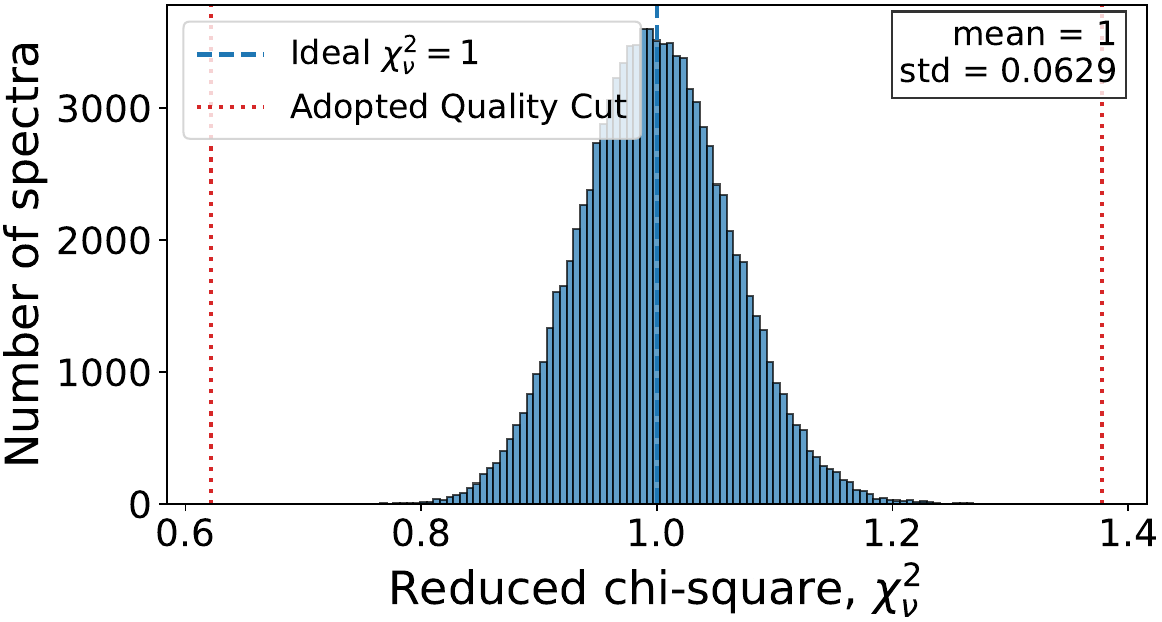}
    \caption{Reduced chi-square distribution obtained from 100,000 synthetic Erlang-distributed spectra. The adopted quality cuts are shown by the vertical dotted lines.}
    \label{fig:synthetic_chi2_reduced}
\end{figure}

To establish the expected variation of the goodness-of-fit statistic for noise-only spectra, we generated 100,000 synthetic spectra with independently exponentially distributed powers. Each was analysed using the same procedure as the experimental data. The resulting distribution of reduced $\chi^2$, shown in Fig.~\ref{fig:synthetic_chi2_reduced}, had a mean and standard deviation of 1.00 and 0.0629, respectively. These values were used to set the data-quality acceptance criteria:  $0.622 < \chi^2_\nu < 1.378$, corresponding to approximately $1\pm6\sigma_{\chi_\nu^2}$. The cut rejected two spectra from the first pass and two from the second-pass out of the eligible tuning folders. The first failures exhibited anomalous background behaviour, while the second failures had substantially elevated $\chi^2_\nu$ values and resembled corrupted start-up records.

\subsection{Averaging}
Within each dataset and cavity tuning, the accepted time-ordered spectra were partitioned into consecutive, non-overlapping groups containing at most 100 spectra, with group sizes made as uniform as possible. This provides an enhancement of signal-to-noise ratio of each group, whilst the conservative probability of a doppler drift of one bin remains low ($\sim1$ bin width drift per 1000 spectra). Each group was represented by the bin-wise arithmetic mean of its baseline-normalised spectra, maintaining the actual group size $N_i$, which was used in subsequent statistical calculations. This procedure produced $17,421$ and $4361$ first and second pass averaged groups, respectively, prior to the subsequent calibration quality and interference vetoes.

\subsection{Native-bin correlations}
\label{sec:correlations}
The raw FFT powers exhibit correlations between neighbouring native frequency bins. For each dataset, the Pearson correlation coefficient $\rho_{l}$ between bins was evaluated before MS filtering using 4420 first-pass and 1370 second-pass spectra, and the median was taken across spectra. The correlations are dominated by the first three lags, with $\rho_1\approx 0.8$ for both datasets. The representative FFT oversampling factor was obtained by jointly fitting the first twelve correlation lags of the first and second-pass datasets to the rectangular-record model derived in Appendix.~\ref{app:fft_response}.

Let $q$ denote the number of stored frequency bins in the sliding search window. For the mean $\bar{x}_q=q^{-1}\sum_{j=1}^{q}x_j$, assuming locally stationary bin statistics with a variance $\sigma_x^2$, summing the covariance terms gives \cite{flyvbjerg1989error} \begin{equation}
    \mathrm{Var}(\bar{x}_q)=\frac{\sigma_x^2}{q^2}(q+2\sum_{l=1}^{q-1}(q-l)\rho_l)=\frac{\sigma_x^2}{q}V_q
\end{equation} where \begin{equation}
    V_q=1+\frac{2}{q}\sum_{l=1}^{q-1}(q-l)\rho_l.
    \label{variance_inflation}
\end{equation} Here $l$ is the integer separation between bins. Averaging $N_i$ independent spectra therefore gives variance $\frac{V_q}{N_i q}$. Equating this to the variance $\frac{1}{k}$ of a unit-mean gamma distribution yields an effective shape parameter $k_{eff,i}=\frac{N_i q}{V_q}$. For the widest search window, $q=256$, the measured values are $V_q\approx 4.1$ for both datasets. These correlations were used in subsequent trigger-threshold and sensitivity calculations.

\subsection{Interference Masks and Vetoes}
\label{section:cuts}
Four classes of contaminated intermediate-frequency (IF) intervals were catalogued: tuning-specific strong instrumental features, dataset-wide IF lines, tuning-specific radio-frequency interference (RFI), and spectral combs, summarised in Table.~\ref{tab:interference_vetoes}.
\begin{table*}[t]
\centering
\caption{Interference vetoes across the data.}
\label{tab:interference_vetoes}
\begin{tabular}{lcl}
\hline
Veto class & Count & Reason/Description \\
\hline
Strong instrumental features & 11 &
\shortstack[l]{Linear combinations of the pump, synth, and/or sampling frequencies result in measured signals.} \\[6pt]

Dataset-wide IF lines & 10 &
\shortstack[l]{Approximately fixed IF across cavity tunings,\\
indicating receiver-chain interference.} \\[6pt]

Tuning-specific RFI & 2 &
\shortstack[l]{Discontinuous frequency jumps over time, consistent with instrumental pickup.} \\[6pt]

Spectral combs & 7 &
\shortstack[l]{Periodic instrumental interference combs with $99\,999.5$~Hz spacing.} \\
\hline
\end{tabular}
\end{table*}

Each tuning-specific RFI signal was carefully analysed over the duration at which they appeared. The key indication that these were pickup of an instrumental source elsewhere was inferred from its stability on the order of minutes, followed by discontinuous frequency jumps up to 100 Hz.

The strong feature masks were applied before baseline estimation to prevent prominent pump and mixing products from distorting the MS baseline.

For a $q$-bin search, each interval was expanded on either side by $(q-1) \frac{\delta f}{2}$, where $\delta f$ is the native bin spacing, so that any sliding window intersecting a masked bin was vetoed. All manual, IF, RFI, and comb vetoes were applied before persistence matching, and consequently, vetoed triggers could neither seed nor support a candidate. Additionally, affected groups provide no coverage of that trial frequency.

\subsection{Multibin trigger catalogues}
Each averaged spectrum was searched using bin-aligned top-hat templates formed from sliding means of $q$ adjacent native bins, evaluated with a one-bin stride. The tested window sizes were $q\in \{ 1,2,4,8,16,32,64,128,256 \}$, corresponding to bandwidths starting from the native bin spacing $\delta f=0.465661$, up to $119.209$~Hz.

For the averaged group $i$, a window was recorded as a trigger when its mean power exceeded \begin{equation}
    x_{\mathrm{th},i}(q)=\Gamma_{\mathrm{sf}}^{-1} \left(e^{-12};k_{\mathrm{eff},i}(q),\frac{1}{k_{\mathrm{eff},i}(q)}\right),
    \label{eq:inverse_survival}
\end{equation} where $\Gamma_{\mathrm{sf},i}^{-1}$ is the inverse survival function of the gamma distribution, and $k_{\mathrm{eff},i}$ is the effective shape parameter defined in Sec.~\ref{sec:correlations}. The value $e^{-12}$ provides a manageable number of triggers for processing ($\sim e^{-12} \times 3.4\times 10^6\approx 21$ per spectra), whose IF frequency and sliding mean power were stored.

\section{Candidate Selection}
An axion flow would produce a narrow excess of microwave power above the receiver-noise background, where the intrinsic linewidth is set by the flow's velocity dispersion. For the persistent flows we consider here, a dark matter signal should recur in observations covering the same radio-frequency range within allowed relative Doppler drifts of the detector. Candidate signals are therefore identified by testing whether an excess is persistent across temporally-separated measurements within the allowed frequency drift. Candidate selection was performed separately for each search width using the combined first and second-pass trigger catalogues. Only the groups and triggers passing the calibration and interference selections discussed earlier enter this stage.

\subsection{Doppler-Drift Model}
The axion signal present at the detector may drift in frequency due to model-dependent Doppler effects. The axion energy depends on the relative velocity $\vec{v}=\vec{v}_{a}-\vec{v}_{D},$ where $\vec{v}_{a}$ is the axion flow velocity, and $\vec{v}_{D}$ is the detector velocity, using the galactic coordinate system. The detector velocity is broken up into several components: $\vec{v}_D =\vec{v}_{\odot} + \vec{v}_e + \vec{v}_r.$ $\vec{v}_{\odot}$ is the Sun's total galactocentric velocity, including the contributions of local standard of rest and the solar peculiar orbit, $\vec{v}_e$ is the velocity of the Earth around the sun, and $\vec{v}_r$ is the velocity of the experiment due to rotation of the Earth in Perth, Australia. The axion frequency is given by Eq.~\ref{eq:axion_energy}. One can treat $\vec{v}_a$ and $\vec{v}_{\odot}$ as constant over the duration of the experiment, and define the axion rest frequency $\nu_{a,0}=\frac{m_a c^2}{h}$. For two spectra taken at $t_1$ and $t_2$, the predicted frequency change is \begin{equation}
    \Delta \nu_a = \frac{\nu_{a,0}}{2 c^2}(\vec{v}(t_2)\cdot \vec{v}(t_2)-\vec{v}(t_1) \cdot \vec{v}(t_1)),
    \label{fdrift}
\end{equation} highlighting the time dependence of the relative velocity $\vec{v}(t)$.

The detector velocity was evaluated by constructing the laboratory barycentric state and transforming it into the Astropy Galactocentric frame. The Earth's barycentric position and velocity at each spectrum time were obtained from the
JPL DE440 ephemeris \cite{park2021jpl}, while the laboratory topocentric velocity due to the Earth's rotation was calculated at the UWA laboratory position using Astropy's \textit{EarthLocation.get\_gcrs\_posvel} routine \cite{2022ApJ...935..167A}. The topocentric GCRS vector is geocentric but oriented consistently with the barycentric celestial axes, so it was combined with the Earth's barycentric state before transformation to the Galactocentric frame. The Galactocentric transformation used the Astropy default solar parameters, corresponding to \(\vec{v}_{\odot}=(12.9,245.6,7.78)\,\mathrm{km\,s^{-1}}\)
\cite{2022ApJ...935..167A}. The resulting detector velocity \(\vec{v}_{D}(t)\) was used with an assumed axion-flow velocity \(\vec{v}_{a}\) to compute the relative velocity \(\vec{v}(t)=\vec{v}_{a}-\vec{v}_{D}(t)\), which determines the expected frequency drift between spectra through Eq.~\ref{fdrift}.

The axion flow speed was fixed at the galactic escape velocity $|\vec{v_{a}}|=v_{esc}=550$ km/s \cite{koppelman2021determination,necib2022substructure} to allow maximal Doppler drift. Its direction was chosen from the two directions parallel and antiparallel to $\vec{v}_D(t_{\rm end})-\vec{v}_D(t_{\rm start})$, retaining the direction that maximised the absolute frequency change between the first and last data timestamps of the experimental run. For a seed trigger at RF frequency $f_s$ in group $i$ and a comparison group $j$, Eq.~\ref{fdrift} was evaluated between the start of the earlier group and the end of the later group, using $f_s$ in place of $\nu_{a,0}$. A trigger at frequency $f_j$ was considered persistent when \begin{equation}
\left|f_j-f_s\right|\leq\left|\Delta f_{ij}\right|+\delta f,
\label{eq:drift}
\end{equation} where $\Delta f_{ij}$ is the corresponding Doppler displacement and $\delta f$ is the native bin spacing.
\subsection{Persistence Requirement}
\label{sec:persistence}
For each seed trigger $s$, $N_{\mathrm{scan},s}$ was the number of eligible averaged groups whose unvetoed RF span contained $f_s$. The corresponding $N_{\mathrm{match},s}$ counted groups containing at least one same-width trigger satisfying Eq.~\ref{eq:drift}. Each group, including the seed group, contributed at most once. The persistence ratio was therefore \begin{equation}
    R_{\mathrm{pers},s}=\frac{N_{\mathrm{match},s}}{N_{\mathrm{scan},s}}\geq0.3.
    \label{eq:persistence}
\end{equation} This is a conservative estimation under the assumption that we expect an incoming axion signal to be persistent at the same frequency within an allowable Doppler drift.

\section{Sensitivity and Density-Limit Construction}
\label{Sec:sensitivity_and_limits}
\subsection{Signal and cavity response}
For this high-resolution analysis, we constrain the axion-flow density $\rho_a$ with the axion--photon coupling fixed to the KSVZ benchmark. Using the normalisation adopted in Ref.~\cite{PhysRevD.96.123008}, the coupling magnitude is \begin{equation}
    g_{\mathrm{KSVZ}}(f)=\frac{0.97 \alpha}{\pi} \frac{hf}{\Lambda^2},
\end{equation} where $\alpha$ is the fine-structure constant and $\Lambda=\SI{0.0776}{GeV}$ is the adopted quantum chromodynamics parameter. The axion rest energy is approximated by $h f$, giving the coupling in GeV$^{-1}$.

For group $i$, the expected axion-conversion power at frequency $f$ and trial flow density $\rho_a$ is \cite{PhysRevD.96.123008,ALKENANY201711} \begin{equation}
\begin{aligned}
    P_{a,i}(f;\rho_a)= &(g_{\mathrm{KSVZ}}^{(J)})^2\frac{\rho_a^{(SI)} \hbar^3 c^3}{\mu_0 (hf)^2}
    \\
    &\times (2\pi f_{c,i}) B_{0}^2 V C_i Q_{L,i} \frac{\beta_i}{1+\beta_i}\mathcal{L}_i(f),
\end{aligned}
\end{equation} where $g_{\mathrm{KSVZ}}^{(J)}$ is the benchmark coupling evaluated at $f$ in units of $J^{-1}$, and $\rho_a^{(SI)}$ is the flow density expressed in $\SI{}{J m^{-3}}$. The Lorentzian cavity response is \begin{equation}
    \mathcal{L}_i (f) = \frac{1}{1+[2(f-f_{c,i})/\Delta f_{c,i}]^2}.
\end{equation} The remaining parameters are the fitted cavity frequency $f_{c,i}$, cavity full width at half max $\Delta f_{c,i}$, magnetic field $B_0$, cavity volume $V$, interpolated form factor $C_i=C(f_{c,i}),$ loaded quality factor $Q_{L,i},$ antenna coupling $\beta_i$, $\hbar=h / (2\pi)$, and vacuum permeability $\mu_0.$

The signal is modelled by an intrinsic top-hat power spectrum of width $W=q\delta f,$ centred on the $q$-bin search window. For the assumed top-hat signals, we define the effective axion quality factor as \(Q_a\equiv\nu_a/W\), where \(W\) is the full intrinsic linewidth. The finite record duration broadens this spectrum according to the rectangular-window FFT response. Its expected contribution to the baseline-normalised statistic is \begin{equation}
    s_i(f,q;\rho_a)=\frac{\eta_{\mathrm{MS}}(q)F_q(R)P_{a,i}(f;\rho_a)}{k_B T_{\mathrm{sys},i}(f) q \delta f}.
\end{equation} Here $k_B$ is Boltzmann's constant, $T_{\mathrm{sys},i}$ is the system-noise temperature, $\eta_{\mathrm{MS}}(q)$ is the MS filter efficiency, and $F_q(R)$ with argument of oversampling factor $R$, is the dimensionless signal-response correction due to oversampling that is derived in Appendix.~\ref{app:fft_response}. The quoted signal widths refer to intrinsic linewidths before instrumental broadening. The limits assume that the signal is centred on the search window.

\subsection{Modified-sinc filter attenuation}
\label{Sec:filter_attenuation}
For each $q$, a weak, unit-area, aligned top-hat was added to a flat baseline, the MS normalisation was applied, and the maximum recovered $q$-bin sliding sum was compared with the injected sum. The resulting efficiencies are shown in Table~\ref{tab:ms_attenuation}.

\begin{table}[htbp]
\centering
\caption{Modified-sinc filter signal-recovery efficiency for each aligned top-hat search width. The efficiency $\eta_{\mathrm{MS}}$ is the ratio of recovered to injected $q$-bin signal power after baseline normalisation.}
\label{tab:ms_attenuation}
\begin{tabular}{
    S[table-format=3.0]
    S[table-format=3.3]
    S[table-format=1.6]
}
\toprule
{$q$} & {$q\delta f$ (\si{\hertz})} & {$\eta_{\mathrm{MS}}$} \\
\midrule
1   & 0.466   & 0.999960 \\
2   & 0.931   & 0.999920 \\
4   & 1.863   & 0.999840 \\
8   & 3.725   & 0.999680 \\
16  & 7.451   & 0.999360 \\
32  & 14.901  & 0.998721 \\
64  & 29.802  & 0.997441 \\
128 & 59.605  & 0.994883 \\
256 & 119.209 & 0.989766 \\
\bottomrule
\end{tabular}
\end{table}

\subsection{Detection probability and 95\% limit}
Using the correlation-corrected shape parameter $k_{\mathrm{eff},i}$ and threshold $x_{\mathrm{th},i}$ defined in Section~\ref{sec:pipeline}, the pipeline represents the signal as an additive shift of the normalised statistic. The probability that a group $i$ crosses its threshold is \begin{equation}
    p_i(\rho_{a};f,q)=\Gamma_{\mathrm{sf}}\left( x_{\mathrm{th},i}-s_i(f,q;\rho_{a});k_{\mathrm{eff},i},\frac{1}{k_{\mathrm{eff},i}} \right),
\end{equation} where $\Gamma_{\mathrm{sf}}$ is   survival function of a gamma distribution. At each trial frequency $f$, and search width $q$, let $N_{\mathrm{scan}}$ denote the number of eligible averaged groups covering that frequency. As in Ref.~\cite{hipp2025search}, groups are treated as independent Bernoulli trials with a trigger probability $p_i$, and thus, the total number of triggering groups, $M$, follows a Poisson-binomial distribution, calculated by successively convolving pairs $(1-p_i,p_i)$. 

For each $f$ and $q$, the $95\%$ KSVZ flow-density limit $\rho_{95}(f,q)$ is obtained numerically by solving \begin{equation}
    \Pr (M\geq n_{\mathrm{req}}|\rho_a=\rho_{95})=0.95,
\end{equation} where $n_{\mathrm{req}}=[0.3N_{\mathrm{scan}}]$, rounded upwards, accounts for the $30\%$ persistence requirement.

\section{Result and Discussion}
\subsection{Data-quality and Candidate Outcomes}
The averaged-group data quality cuts are summarised in Table~\ref{tab:group-selection}. Missing-metadata (No meta.) groups correspond to tuning folders that were not included in the eligible medium-resolution analysis and are therefore not used in the high-resolution search. Averaged-groups with suspicious $T_{\mathrm{sys},i}$ values, and missing spectrum analyser files (No cal.) are also excluded (see Sec.~\ref{sec:tsys}).

\begin{table}[htbp]
\centering
\caption{Averaged-group calibration status before frequency-dependent interference cuts.}
\label{tab:group-selection}
\begin{tabular}{lrrrrr}
\toprule
Data set & Accepted & Bad \(T_{\rm sys}\) & No meta. & No cal. & total\\
\midrule
First & 14,256 & 505 & 2,631 & 29 & 17,421\\
Second   & 3,227  & 850 & 268   & 16 & 4,361\\
\midrule
Combined & 17,483 & 1,355 & 2,899 & 45 & 21,782\\
\bottomrule
\end{tabular}
\end{table}

No triggers survive the $30\%$ persistence requirement after calibration and interference vetoes.

\subsection{KSVZ flow-density limits}
\begin{figure*}[htbp]
    \centering
    \includegraphics[width=0.95\linewidth]{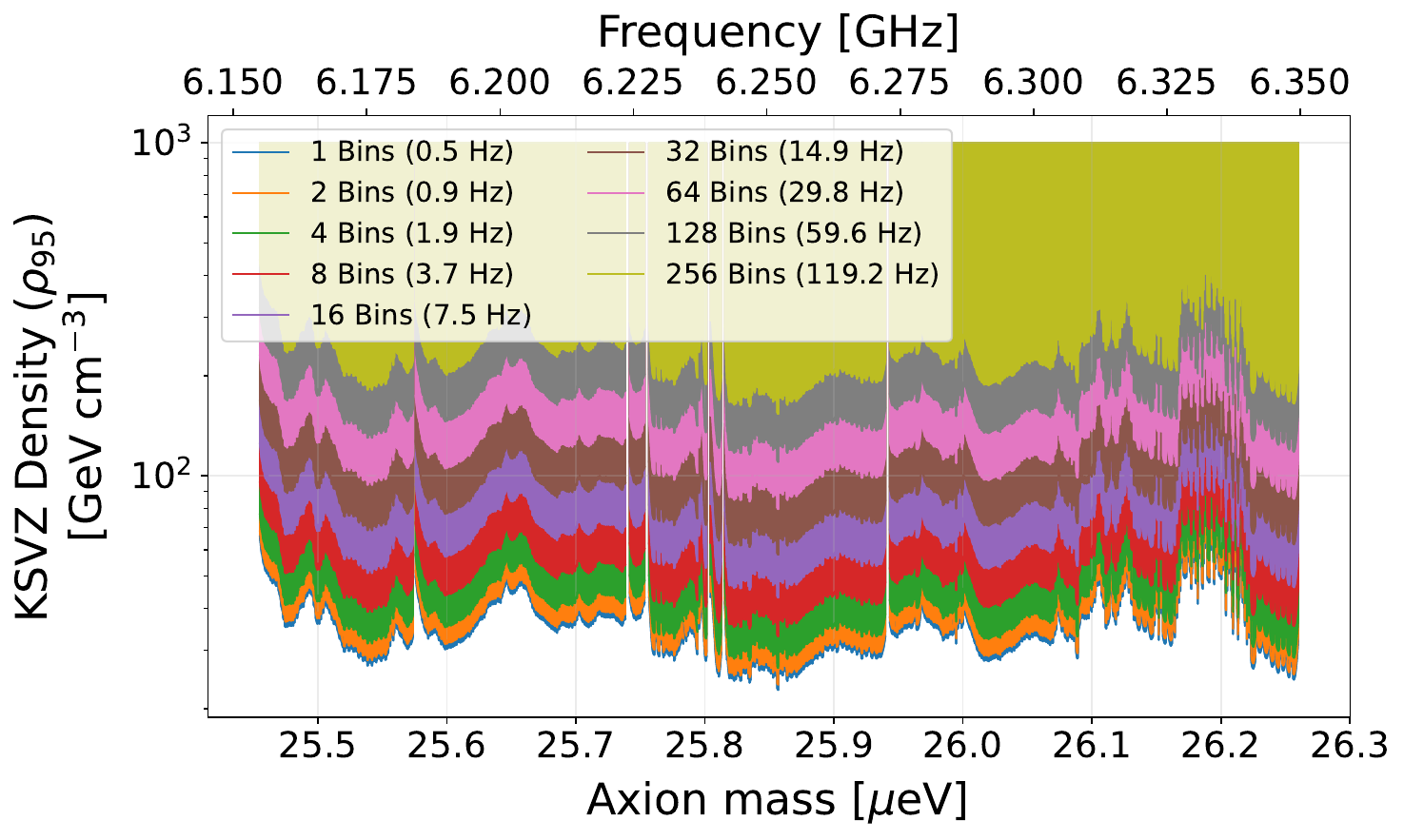}
    \caption{95\% KSVZ axion-flow density limits sampled on a \SI{1}{kHz} grid for the nine aligned top-hat signal widths. At evaluated frequencies, densities above each curve are excluded under the intrinsic top-hat model. The calculation includes MS filter attenuation and the measured native-bin correlation correction.}
    \label{fig:density-limits}
\end{figure*}

Figure~\ref{fig:density-limits} presents the nominal $95\%$ KSVZ axion-flow density limits sampled every \SI{1}{kHz}. Each coloured curve corresponds to a distinct assumed top-hat linewidth. At evaluated points, densities above a curve are excluded for that linewidth; missing points denote no valid post-cut coverage.

The strongest $95\%$ limits placed on KSVZ axion-flow density range from \SI{22.8}{GeVcm^{-3}} for a 0.466 Hz signal to \SI{153.7}{GeVcm^{-3}} for a 119.209 Hz signal.

The effect of strong nearby bin correlations on the order of $\sim 4$ bins manifests as reduced sensitivity particularly to the narrowest few-bin-width searches. 

\subsection{Uncertainty Analysis}
Following the uncertainty prescription of Ref.~\cite{quiskamp2025near}, we adopt fractional uncertainties of $11.6\%$ in $T_{\mathrm{sys}}$ and $5\%$ in $C_i$.
Assuming independence, and holding other experimental parameters as fixed, the signal statistic scales as $\rho_a C/T_{\mathrm{sys}}$. Consequently, the density limit scales as $\rho_{95}\propto T_{\mathrm{sys}}/C$. First-order propagation gives \begin{equation}
    \frac{\delta \rho_{95}}{\rho_{95}} \approx \sqrt{(\frac{\delta T_{\mathrm{sys}}}{T_{\mathrm{sys}}})^2+(\frac{\delta C_i}{C_i})^2}\approx12.6\%.
\end{equation}

\subsection{Interpretation and Comparison}
The present search complements the previous ORGAN-Q virialised-halo analysis \cite{quiskamp2025near} by testing substantially narrower spectral features within the same mass region. The earlier analysis constrained the axion--photon coupling for an assumed local density. Here, we fix the coupling to the mass-dependent KSVZ benchmark and constrain the density of a persistent axion flow under the signal model described in Sec.~\ref{Sec:sensitivity_and_limits}.

Previous ADMX searches placed limits on nonvirialised axion populations in the $\SIrange{1.98}{2.17}{\mu eV}$ and $\SIrange{3.3}{3.69}{\mu eV}$ mass regions, considering different flow velocity dispersions and treatments of Doppler modulation \cite{PhysRevD.74.012006,PhysRevD.84.121302,PhysRevD.94.082001}. More recently, \citeauthor{PhysRevD.109.083014} reported frequency-dependent KSVZ flow-density limits of approximately $\SIrange{0.003}{0.01}{GeV/cm^{3}}$ at $95\%$ confidence across the searched portions of $\SIrange{2.81}{3.31}{\mu eV}$ \cite{PhysRevD.109.083014}. \citeauthor{hipp2025search} extended the coverage to $\SIrange{3.3}{4.2}{\mu eV}$, reporting KSVZ flow-density limits of approximately $\SI{0.004}{GeV/cm^{3}}$ in their finest-resolution analysis at $95\%$ confidence. CAPP-12 TB also searched for axions associated with the Sagittarius tidal stream, excluding KSVZ axion densities above approximately $\SI{0.025}{GeV/cm^{3}}$ over $\SIrange{4.51}{4.59}{\mu eV}$ at $90\%$ confidence under the adopted stream model \cite{PhysRevD.108.L021304}.

The present work provides complementary constraints at substantially higher axion masses, $\SIrange{25.455}{26.260}{\mu eV}$, for nine assumed linewidths between 0.466 Hz and 119.209 Hz. Comparisons between these searches must account for their different signal shapes, velocity dispersions, Doppler treatments, and confidence levels. The benchmark of the ``Big Flow'' discussed in Sec.~\ref{sec:intro} is not excluded, with our strongest density limits presented in this study approximately twice this value. This limit is also roughly 51 times larger than the conventional local-halo density benchmark of \SI{0.45}{GeV/cm^3} \cite{read2014local}.

\section{Conclusion}
The ORGAN-Q high-resolution data were analysed for persistent narrow axion signals over the mass range $\SIrange{25.455}{26.260}{\mu eV},$ using nine assumed linewidths between \SI{0.466}{Hz} and \SI{119.209}{Hz}. After calibration and interference vetoes, no candidate satisfies the $30\%$ persistence requirement. Under the assumed signal model and KSVZ axion--photon coupling, we obtain $95\%$ upper limits on the local axion-flow density. The strongest limits are \SI{22.8}{GeV/cm^{3}} for a 0.466~Hz linewidth, and \SI{153.7}{GeV/cm^{3}} for a 119.209 Hz linewidth, both attained near an axion mass of $\SI{25.857}{\mu e V}$.

Future ORGAN experiments will have a better understanding of the acquisition and analysis pipeline, maturing the pipeline to scout for signs of correlations earlier and improving the efficiency of data processing. Additionally, acquisition procedures that minimise the total data stored will be investigated, aiming to implement real-time filtering and threshold triggers.

By probing spectral features substantially narrower than those expected from a virialised halo, this analysis extends the ORGAN-Q experiment to non-virialised axion populations. These density constraints complement the previously reported virialised-halo coupling limits within the same mass range.

\begin{acknowledgments}
This work is supported by the Australian Research Council grant CE200100008. 
Michael T. Hatzon acknowledges support from the Defence Science Centre, an initiative of the State Government of Western Australia, and from The University of Western Australia through the Hackett Postgraduate Research Scholarship. Ben T. McAllister acknowledges support from grant DE250100933.
\end{acknowledgments}

\appendix 
\section{FFT Response due to Oversampling}
\label{app:fft_response}
For white Gaussian noise sampled by a rectangular window, the correlation between nearby values due to oversampling follows the squared-sinc form derived in Appendix A of Ref.~\cite{10.1093/mnras/stv288}. This is given by \begin{equation}
    \rho_{\ell} \approx \operatorname{sinc}^2(T_{\rm rec} \ell \delta f)=\operatorname{sinc}^2(\frac{\ell}{R}),
\end{equation} where $\ell$ is the separation in stored bins, $R=(T_{\rm rec}\delta f)^{-1}$, and $\operatorname{sinc}(x)=\frac{\sin(\pi x)}{\pi x}.$ 

For a rectangular window of duration $T_{\rm rec}$, the unit-area spectral response to a monochromatic signal, evaluated at a frequency offset $\Delta f$ from the signal, is \begin{equation}
\begin{aligned}
    K(\Delta f) &= \frac{1}{T_{\rm rec}} \left| \int_0^{T_{\rm rec}}e^{-2 \pi i \Delta f t}dt \right|^2\\
    &=T_{\rm rec} \operatorname{sinc}^2(T_{\rm rec} \Delta f).
\end{aligned}
\label{eq:fft-response-kernel}
\end{equation}
We note here that $\int_{-\infty}^\infty K(\Delta f)d \Delta f=1,$ therefore a monochromatic signal of total power $P$ has a measured signal PSD of $PK(\Delta f).$

Next, we consider the top-hat signal, described by an intrinsic power spectrum that is uniform over the full width $W$, with intrinsic PSD \begin{equation}
    S_{\rm intrinsic}(\nu_a + \nu')= P L_W(\nu').
\end{equation} Here $\nu_a$ is the central signal frequency, $\nu'$ is the frequency offset from $\nu_a$, and $L_W(\nu')=1/W$ for $|\nu'|\leq W/2,$ and zero otherwise. A frequency interval $d\nu'$ within this band contains mean signal power, \begin{equation}
    dP=\frac{P}{W}d\nu'.
\end{equation} At the evaluation frequency $f=\nu_a+\Delta f$, the contribution from a component at $\nu_a +\nu'$ is $K(\Delta f-\nu')dP.$ For the assumed stationary signal with an intrinsic top-hat PSD, the expected measured signal PSD is obtained by convolving the intrinsic PSD with the sampled response $K$: \begin{equation}
\begin{aligned}
    S_{\rm sig, T}(f) &=\frac{P}{W}\int_{-W/2}^{W/2}K(\Delta f -\nu') d\nu',\\
    &=P H_W(\Delta f),
    \end{aligned}
    \label{eq:top-hat-response}
\end{equation} where $H_W(\Delta f)$ is the top-hat shape after the instrumental response, with contributions added in mean power.

For a signal centred on a $q$-bin window, the frequency bin offsets are \begin{equation}
    \Delta f_j = (j-\frac{q-1}{2})\delta f, \quad j=0,...,q-1.
\end{equation} Denoting the local noise PSD as $S_0=k_B T_{\rm sys}$ for Boltzmann's constant $k_B$ and system temperature $T_{\rm sys}$, before any filter attenuation the expected excess above the unit baseline of the normalised $q$-bin mean is \begin{equation}
\begin{aligned}
    s_q&=\frac{P}{q S_0}\sum_{j=0}^{q-1}H_W(\Delta f_j)\\
    &=\frac{P F_q}{q S_0 \delta f},
\end{aligned}
\end{equation} which defines the dimensionless response factor \begin{equation}
    F_q=\delta f \sum _{j=0}^{q-1}H_W(\Delta f_j).
    \label{eq:fft-factor-definition}
\end{equation}

For the adopted intrinsic width $W=q \delta f$, define \begin{equation}
    u_j = \frac{j-\frac{q-1}{2}}{R}, \quad a_q=\frac{q}{2R}.
\end{equation} Substituting Eqs.~\ref{eq:fft-response-kernel} and \ref{eq:top-hat-response} into Eq.~\ref{eq:fft-factor-definition}, and changing the integration variables gives \begin{equation}
   F_q(R)= -\frac{1}{q}\sum_{j=0}^{q-1} \int_{u_j+a_q}^{u_j-a_q} \operatorname{sinc}^2(u)du.
\end{equation} Evaluating the integral where \begin{equation}
    J(u)=\frac{\operatorname{Si}(2 \pi u)}{\pi}-u \operatorname{sinc}^2(u),
\end{equation} is an anti-derivative of $sinc^2(u)$, and \begin{equation}
    \operatorname{Si}(u)=\int_0^u\frac{\sin t}{t} dt,
\end{equation} yields the final form of the implemented response factor: \begin{equation}
    F_q=\frac{1}{q}\sum_{j=0}^{q-1}[J(u_j+a_q)-J(u_j-a_q)].
\end{equation} For the measured value of $R$, representative factors are $F_1\approx 0.23$, $F_4 \approx 0.64$, and $F_{256}\approx 0.99$.

At fixed axion--photon coupling, the conversion power is proportional to the flow density. Consequently, a smaller signal response requires a larger density to produce the same expected excess.

The combined signal response is modelled by the factorised approximation $\eta_{MS}(q) F_q(R)$, using MS filter efficiencies obtained for unbroadened top-hat profiles. The correlated-noise variance is treated specifically using the measured frequency-bin correlations.

\bibliographystyle{apsrev4-2}
\bibliography{hrbib}
\end{document}